\documentstyle[12pt]{article}
\newcommand{\bea}{\begin{equation}}
 \newcommand{\eea}{\end{equation}}
 \newcommand{\ber}{\begin{eqnarray}}
 \newcommand{\eer}{\end{eqnarray}}
\begin{document}
%\begin{center}
\title{\bf Cosmic Optical Activity in a Randall-Sundrum Braneworld with Bulk Kalb-Ramond Field}
\author{Debaprasad Maity\footnote{E-mail: tpdm@iacs.res.in}~ and~ 
Soumitra SenGupta\footnote{E-mail: tpssg@iacs.res.in}\\
Department of Theoretical Physics\\
Indian Association for the Cultivation of Science\\
Jadavpur, Kolkata 700 032, India}
\date{}
\maketitle
%\end{center}

PACS number(s): 04.50.+h, 11.25.Wx, 04.80.Cc

\begin{abstract}
Optical activity of electromagnetic waves in a string inspired Kalb-Ramond
cosmological background 
is studied in presence of extra spacetime dimension.
The Kalb-Ramond-electromagnetic coupling which originates from the gauge anomaly
cancelling Chern-Simons term in a string inspired model, is explicitly calculated following Randall-Sundrum
braneworld conjecture.
It is shown that the Randall-Sundrum scenario
leads to an enormous enhancement of the optical rotation of a plane polarized electromagnetic
wave propagating on the visible brane.Absence of any  experimental support in favour of such a large
rotation in astrophysical experiments on distant galactic 
radio waves indicates an apparent conflict between Randall-Sundrum
brane world scenario and the presence of Kalb-Ramond antisymmetric tensor field in the background spacetime.
\end{abstract}

\thispagestyle{empty}

\newpage
\section{Introduction}
Possibility of observing some experimental signature of String theory in the present 
low energy world is a subject of interest for a long time.One of the important testing arena
is considered to be the Astrophysical/Cosmological observations.
Various massless modes of string theory, which are obviously most relevant for the low energy
world are expected to have some new observable effects on cosmic phenomena. 
Among the various massless modes, the low energy field theory action of closed String theory 
has a second rank antisymmetric tensor field known as Kalb-Ramond (KR) field $B_{\mu\nu}$.
The corrresponding KR fieldstrength
$H_{\mu\nu\lambda}=\partial_{[\mu}B_ {\nu\lambda]}$ is modified by 
Chern-Simons terms which originate from the requirement of 
quantum consistency namely, the gauge anomaly and gravitational anomaly 
cancellation in the underlying string theory. 
The Chern-Simons term for the $U(1)$ gauge anomaly cancellation has been shown to play crucial role in preserving 
the $U(1)$ gauge symmetry in the resulting action and thereby providing with a gauge invariant coupling
between the KR and the electromagnetic field \cite{pmssg}. The resulting gauge invariant coupling of the
KR field to the Maxwell field allows us to study the phenomenological effects of string theory on the 
propagation of electromagnetic field in a KR background\cite{skssgss,ssgss,skpmssg,cqg,helflip}.
It turns out that this leads to
the optical activity of the electromagnetic wave
passing through such a space-time. In four dimension,this effect of
optical activity has already been explored \cite{skpmssg,cqg} with different cases of
space,time dependence of the pseudoscalar field H (dual of the
massless three form $H_{\mu\nu\lambda}$). However experimental bounds on the 
optical activity implies  that
the H
field must be very weak so that it's contribution to the obeserved activity
in addition to the usual Faraday rotation must be very low \cite{nodland,wardle}. 

As a possible explanation to this result
it was shown that the effect of extra dimension could be a
possible reason for the suppression
of the KR field although the pure gravity sector and the KR field has
identical coupling in the pre-compactification scale.
KR field being the massless mode of a closed string is taken on the bulk alongwith the gravity.
It has been explicitly shown \cite{bmssg} that in a 
higher dimensional framework of Randall-Sundrum
\cite{RS}scenario the massless mode of KR field in the visible
brane is suppressed by a large exponential warp factor. This motivates us to
explore whether the phenomenon of optical activity also suffers large suppression 
in a higher dimensional scenario. For this one needs to study the effective coupling between the
electromagnetic field and KR field ( arising from Chern-Simons extension) after compactification.

     In a string inspired model both the gravity and the KR field are massless modes of closed 
strings and therefore are assumed to propagate in the 
bulk while all the standard model fields are confined on  the visible 3-brane.In a subsequent section we 
shall take up a more general case where the  U(1) electromagnetic field also propagates in the bulk.
Possibility of such a scenario in the context of a large internal dimension was considered earlier
to explore various aspects of supersymmetry breaking \cite{antonio}. Later it was generalized for the braneworld model.
There are various reasons to consider gauge field entering into the bulk.To understand the geometric origin
of the spectrum of the fermion masses\cite{msch,mashif,kaplan},to identify the Higgs as the extra-dimensional
component of the gauge fields(to protect it's mass from correction)\cite{ljhall},to provide a viable candidate
for dark matter\cite{gservant},to achieve high scale gauge coupling unification\cite{dudas,dienes} and many other  
importent issues led to the model of a bulk $U(1)$ gauge field in a braneworld scenario.  
\section{Optical Activity in a KR background}

In the proposed string inspired model adopted by us,the
higher dimensional extension of the low energy effective action for the gravity
and electromagnetic sectors in 5-dimension is given by,
\begin{equation}
S = \int d^5 x \sqrt{- g} \left[ R(g) - \frac 1 4 F_{M N} F^{M N} + \frac 1 2
\tilde H_{M N L} \tilde H^{M N L} \right]
\end{equation}

where $\tilde H_{M N L} = \partial_{[M} B_{N L]}+\frac 1 {3M_p^{\frac 1 2}}\delta_M^{\mu}\delta_N^{\nu}
\delta_L^{\lambda}A_{[\mu}F_{\nu \lambda]} $
with each Latin index running from 0 to 4 and Greek index running from 0 to 3. 
The second term on the righthand side of this equation is the $U(1)$Chern-Simons
term.\\
Assuming that the  gauge field is confined on the visible 3-brane 
the above low energy effective action for the
gravity and
KR field coupled to the $U(1)$ electromagnetic field reduces to
\ber
 S &=& \int d^5 x \sqrt{- g} \left[ R(G)  + \frac 1
2  H_{M N L}  H^{M N L}\right]-
\frac 1 4\int d^4x \sqrt{-g_{vis}}F_{\mu\nu}F^{\mu\nu}\nonumber\\
 & &+ \frac 1 {3M_p^{\frac 1 2}}\int d^5x \sqrt{-g} H_{M N L}
\delta_{\mu}^M\delta_{\nu}^N\delta_{\lambda}^L A^{[\mu}F^{\nu \lambda]}\delta(\phi-\pi)
\eer
We have ignored the higher order Chern-Simons term because of extra Planck mass
suppression.
It has been shown that  in the case of four dimensional scenario the
KR-electromagnetic field coupling 
leads to the phenomena of optical activity.\\  
In a flat four dimensional
background this rotation of the plane of polarization \cite{skpmssg} in terms
of the comoving time $\eta$ is given as,
\bea 
(\Delta \phi)_{mag}= -h \eta/M_p 
\eea 
$h$ comes from the dual pseudoscalar field $H$ defined through the duality relation
\begin{equation}
H_{\mu\nu\lambda } = \epsilon_{\mu\nu\lambda\sigma} \partial^{\sigma} H
\end{equation}
where the solution for $H$ is given as $H = h \eta + h_{0}$.
We shall like to emphasize here that equ.(2) and equ.(4) together imply an axion-elctromagnetic coupling
which has emerged naturally from the requirement of the gauge anomaly cancellation in the underlying string theory.
Thus such an axion induced effects on electromagnetic field 
is a direct consequence of the requirement of quantum consistency of the full string theory.   

The  generalization of this optical activity with inhomogeneous Kalb-Ramond field in a  non-flat background has also
been done in ref \cite{cqg}.
While this axion induced optical rotation may be viewed as  a possible explanation of the additional
small wavelength independent rotation of the plane of
polarization of the
electromagnetic wave from distant galaxies over the usual Faraday rotation \cite{nodland}, a
comparison with experimental data immediately implies that the pseudoscalar field(dual to the 3 form KR field strength ) $H$ must
couple very weakly to the electromagnetic field.

In another work\cite{bmssg} it has been shown that although in a higher dimensional  scenario the
KR field and gravity have similar coupling
at the Planck scale, a compactification in Randall-Sundrum scenario suppresses the KR field enormously on the 
visible 3-brane.
In that scenario it was assumed that like gravity KR field also resides in the
bulk whereas the standard model fields are
confined on the visible 3-brane.

In this paper we explore whether a RS type of extra dimensional brane world picture 
may lead to the suppression of KR field induced
optical rotation in the brane to a near invisibility.To startwith we take the
elctromagnetic  and the other standard model
fields to be confined in the visible 3-brane whereas the gravity and KR field propagates in
the bulk.Later we shall consider the scenario where the $U(1)$ gauge field also propagates in the bulk.
\begin{center}
\bf{Optical activity in Randall-Sundrum scenario}
\end{center}
In RS scenario in a five dimensional ADS spacetime, the fifth coordinate $\phi$
is compactified on a $S_1/Z_2$ orbifold. Two branes namely the hidden brane and
the visible brane
are located at the two orbifold fixed points $0$ and $\pi$ respectively.
It was shown that the corresponding background metric is given as,

\begin{equation}
ds^2 = e^{-2 \sigma} \eta_{\alpha \beta} dx^{\alpha}  dx^{\beta}  + r_c^2 d\phi^2
\end{equation}
where $\sigma = k r_c  \vert \phi  \vert   $ . Now consider the free 
electromagnetic part of the action in eqn.(2),
\begin{equation}
S_{electro}=- \frac 1 4 \int d^4x \sqrt{-g_{vis}}g^{\mu\alpha}
g^{\nu\beta} F_{\mu\nu} F_{\alpha\beta}
\end{equation}
Assuming that the electromagnetic field to be confined on the flat visible brane and noting that the 
$\sqrt{-g_{vis}}=e^{-4kr_c\pi}$and $g^{\mu\nu}
=e^{2kr_c\pi}$ the above action reduces to
\begin{equation}
S_{electro}=- \int d^4 x \eta^{\mu\alpha}\eta^{\nu\beta}
 F_{\alpha \beta} F_{\mu\nu}
\end{equation}
Similarly the 5-dimensional action
corresponding to the Kalb-Ramond field is given by
\begin{equation}
S_H = \frac 1 2 \int d^5 x \sqrt{-g} H_{M N L}H^{M N L}
\end{equation}
where $\sqrt{-g}=e^{-4 \sigma} r_c$. This action has KR gauge invariance
$\delta B_{MN} = \partial_{[M}\Lambda_{N]}$.

We use the KR gauge fixing condition to set
$ B_{4 \mu}=0$. Therefore the only non vanishing KR field components are $B_{\mu\nu}$ where
$\mu ,\nu$ runs from 0 to 3. 
These components are functions of both compact and
non-compact coordinates. One thus gets
\begin{equation}
S_H = \frac 1 2 \int d^4 x \int d{\phi}  r_c  e^{2 \sigma(\phi)}
\left[\eta^{\mu \alpha} \eta^{\nu \beta} \eta^{\lambda \gamma} H_{\mu \nu
\lambda} H_{\alpha \beta \gamma} - \frac 3 { r^2_c} e^{-2 \sigma(\phi)}
\eta^{\mu \alpha} \eta^{\nu \beta} B_{\mu\nu} \partial^2_\phi B_{\alpha \beta} \right]
\end{equation}
Applying  the Kaluza-Klein decomposition for the Kalb-Ramond field:
\begin{equation}
B_{\mu \nu}(x,\phi) = \sum^{\infty}_{n=0} B^n_{\mu \nu}(x) \chi^n(\phi) \frac 1
{\sqrt{r_c}}
\end{equation}
and demanding that in four dimension an effective action for $B_{\mu \nu }$ should be of the form
\begin{equation}
S_H = \int d^4 x \sum^{\infty}_{n=0} \left[ \eta^{\mu \alpha} \eta^{\nu \beta}
\eta^{\lambda \gamma} H^n_{\mu \nu \lambda} H^n_{\alpha \beta \gamma} + 3 m^2_n
\eta^{\mu \alpha} \eta^{\nu \beta} B^n_{\mu \nu} B^n_{\alpha \beta}\right]
\end{equation}
where $H^n_{\mu \nu \lambda} = \partial_ {[\mu} B^n_{nu \lambda]}$  and $\sqrt
{3} m_n $ gives the mass of the nth KR mode,

one obtains
\begin{equation}
- \frac 1 {r^2_c} \frac {\partial^2 \chi^n} {\partial \phi^2} = m^2_n \chi^n
e^{2 \sigma}
\end{equation}
The $\chi^n(\phi)$ field satisfies the orthogonality condition
\begin{equation}
\int e^{2 \sigma(\phi)} \chi^m(\phi) \chi^n(\phi) d \phi = \delta_{m n}
\end{equation}
Defining $z_n = e^{\sigma(\phi)} m_n/k$ the above equation reduces to
\begin{equation}
\left[ z^2_n \frac {d^2} {d z^2_n} + z_n \frac d {d z_n} + z^2_n \right] \chi^n
= 0
\end{equation}
This has the solution
\begin{equation}
\chi^n = \frac 1 {N_n}[ J_0(z_n) + \alpha_n Y_0(z_n)]
\end{equation}
The zero mode solution \cite{bmssg} of $\chi $ therefore turns out to be 
\begin{equation}
\chi^0 (\phi) = C_1 \vert \phi \vert + C_2
\end{equation}
However the condition of self-adjointness leads to $C_1 = 0$  and leaves the scope of only a constant
solution for $\chi^{0}(\phi)$.Using the normalization condition, one finally obtains
\begin{equation}
\chi^0 = \sqrt {k r_c} e^{-k r_c \pi}
\end{equation}
This result clearly indicates that the massless mode of the KR field is suppressed by a large warp factor on the visible 
3-brane.
In a similar way we now take the KR-EM interaction term 
\begin{equation}
S_{int} = \frac 1 {3M_p^{\frac 1 2}} \int d^5 x \sqrt {-g} 
H_{M N L}\delta_{\mu}^M\delta_{\nu}^N\delta_{\lambda}^L
 A^{[\mu}F^{\nu\lambda]}\delta(\phi - \pi)
\end{equation}
Following the same arguments as were given in the previous case,the interaction
term reduces to 
\begin{equation}
S_{int} = \frac {r_c} {M_p^{\frac 1 2}} \int d^4 x\int d\phi e^{2\sigma}\delta(\phi-\pi)
 \eta^{\mu\alpha}\eta^{\nu\beta}\eta^{\lambda\gamma}
H_{\mu \nu \lambda}(x,\phi)A_{[\alpha}F_{\beta \gamma]}
\end{equation}
Integrating over the bulk coordinate $\phi$ ,  
retaining only the massless modes and using eq.(18) one obtains
\begin{equation}
S_{int} = \sqrt {\frac k M_p}r_c e^{k r_c \pi}  \int d^4 x 
 H_{\mu \nu \lambda}(x) A^{[\mu}F^{\nu \lambda]}
\end{equation}
With these the KR-electromagnetic part of the action becomes
\begin{equation}
S = -\frac 1 4 \int d^4 x F_{\mu \nu} F^{\mu \nu} + \frac 1 2 \int d^4 x H_{\mu
\nu \lambda} H^{\mu \nu \lambda} +\frac 1 3 \sqrt {\frac k M_p}r_c e^{k r_c \pi} \int d^4 x
H_{\mu \nu
\lambda}A^{[\mu}F^{\nu \lambda]}
\end{equation}
It may be noted here that in the string inspired RS model our calculation 
completely determines the KR-Maxwell coupling as is evident from the above action.\\ 
Now varying with respect to $B_{\mu \nu}$ and $A_{\mu}$ we obtain the following
equations
\begin{equation}
\partial_{\alpha} \left[H^{\alpha \beta \gamma} + \sqrt {\frac k 
M_p}r_c e^{k r_c \pi}
A^{[\alpha}F^{\beta \gamma]} \right] = 0
\end{equation}

\begin{equation}
\partial_{\alpha} \left[  F^{\alpha \beta} - 2 \sqrt {\frac k M_p} r_c e^{k r_c \pi}
H^{\alpha \beta \gamma} A_{\gamma} \right]
 = -\sqrt {\frac k M_p}r_c e^{k r_c \pi} H^{\alpha \beta \gamma} F_{\gamma \alpha}
\end{equation}
Replacing the massless three form $H^{\alpha \beta \gamma}$ by using the duality 
relation 
\begin{equation}
H^{\alpha \beta \gamma} = \epsilon^{\alpha\beta\gamma\mu}\partial_{\mu}H
\end{equation}
where H is the dual pseudo scalar axion, we find the modified Maxwell's equations as   
 \ber
\bf \nabla \cdot \bf E &=& 4 \sqrt {\frac k M_p}r_c e^{k r_c \pi} \bf \nabla H \cdot
\bf B\nonumber\\
\partial_{0}\bf E - \bf \nabla \times \bf B &=& 4 \sqrt {\frac k
M_p}r_c e^{k r_c \pi}
[\partial_{0}H \bf B + \bf \nabla H \times \bf E]\nonumber\\
\bf \nabla \cdot \bf B &=& 0\nonumber\\
\partial_{0}\bf B + \bf \nabla \times \bf E &=& 0\nonumber\\
 & & {}
\eer
Now we consider the Bianchi identity of the KR field strength
\begin{equation}
\epsilon^{\mu\nu\lambda\sigma} \partial_{\sigma}H_{\mu\nu\lambda} = 0
\end{equation}
This alongwith equ.(24) immediately implies that the pseudoscalar H satisfies the massless
Klein-Gordon equation $\Box H = 0$. In a flat four dimensional spacetime H is
only a function of comoving time
co-ordinate $\eta$.As a result the  Klein-Gordon equation simply reduces to the $  {d^2
H}/{d \eta^2} = 0$ which has the solution $ H = h {\eta} + h_0$,where $h$ and
$h_0$ are constants.Proceeding along the lines of \cite{skpmssg,carrolfield},we arrive at the equation
\begin{equation}
\frac {d^2 b_{\pm}} {d \eta^2} + (p^2 \mp 4 p \sqrt {\frac k M_p}r_c
 e^{k r_c \pi}
h)b_{\pm} =0
\end{equation}
Where we have decomposed $\bf{B} = \bf{b}(\eta)e^{i p\cdot x}$ and have chosen the
z-direction as the propagation direction of the electromagnetic wave, p being the wave vector.The
circular polarization states are defined by $ b_\pm = b_x \pm i b_y$.
So from equ.(27) the optical activity due to the KR field is given by, 
\bea
(\Delta
\phi)_{mag} \equiv \frac 1 2 (\phi_+ - \phi_-) = -2 \sqrt {\frac k 
M_p}r_c e^{k r_c \pi} h
\eta.
 \eea
Comparing with our previous result of the optical rotation in four dimensional
spacetime
we find that Randall-Sundrum scenario causes an enormous enhancement of the optical rotation
in the visible brane although the field H itself gets suppressed by the warp factor.

\section{Coupling betweeen the bulk U(1) gauge field and the Kalb-Ramond field
in  RS Scenario}      
Let us now focus our attention to 
the RS scenario with the electromagnetic gauge field in the bulk.
In this case the action for a bulk electromagnetic gauge field is
given as \cite{bulkgauge}
\bea
S_{gauge} = - \frac 1 4 \int d^5 x \sqrt {-g} F_{M N}F^{M N}
\eea 
where $\sqrt {-g}= r_c e^{-4 \sigma(\phi)}$.
After RS compactification,
\bea
S_{gauge} = -\frac 1 4 \int d^4 x \int d{\phi}\left[r_c \eta^{\mu
\alpha}\eta^{\nu \beta}F_{\alpha \beta} F_{\mu\nu} - \frac 2
{r_c}\eta^{\mu\alpha}A_{\alpha}\partial^2_{\phi}(e^{-2\sigma}A_{\mu})\right]
\eea 
Next,we consider Kaluza-Klein decomposition for the U(1) gauge field $A_{\mu}$ which
is function of both $x$ and $\phi$: 
\bea
A_{\mu}(x,\phi)=\sum^\infty_{n=0}A_{\mu}(x)\frac {\chi^n(\phi)} {\sqrt {r_c}}
\eea 
In terms of four dimensional field $A_{\mu}(x)$,an effective action
of the form
\bea
 S_{gauge} =- \frac 1 4 \int d^4 x \sum^{\infty}_{n=0}\left[
\eta^{\mu \alpha}\eta^{\nu\beta}F^n_{\alpha\beta}F^n_{\mu\nu} + 2 m^2_n
\eta^{\mu\nu}A^n_{\mu}A^n_{\nu}\right]
 \eea
can be obtained provided 
\bea - \frac 1 {r^2_c} \frac {\partial^2(e^{-2
\sigma}\chi^m)} {\partial \phi^2} = m^2_n \chi^m
\eea
alongwith the orthogonality condition
\bea \int d{\phi} \chi^n(\phi)\chi^m(\phi)=\delta_{n m}
 \eea
where $\sqrt {2} m_n $ gives mass of the nth mode of the gauge field. In this
case we have assumed that $A_4$ is a $Z_2$ odd function of the extra dimension and have 
used the gauge degree of 
freedom to choose  $A_4=0$. The gauge invariant equation $\int d^4 x A_4 = 0$ follows automatically
from $Z_2$ odd condition. We have also used the Lorentz gauge
condition
$\eta^{\mu\nu}\partial_{\mu}A^n_{\nu}= 0$ .\\
Now in terms of $z_n = \frac {m_n} {k} e^{\sigma}$ and $y_n =
e^{-\sigma}\chi^n$, one may recast the above equation in the form 
\bea
\left[z^2_n \frac {d^2} {d z^2_n} + z_n\frac d {d z_n} + (z^2_n -1)\right] y_n
= 0 
\eea 
The above equation admits of the  solution:
\bea
 \chi^n = \frac {e^{\sigma}} {N_n} \left [J_1(z_n) + \alpha_n
Y_1(z_n)\right] 
\eea
This yields the  zero mode solution
\bea
 \chi^0 = \frac 1 {\sqrt {2\pi}}
 \eea
Earlier we got the zero mode solution for the KR field \bea
\xi^0 = \sqrt {k r_c} e^{-k r_c \pi}
 \eea
 Now in this scenario we again consider the interaction term
\bea
 S_{int}= \frac 1 {3M_p^{\frac 3 2}} \int d^5 x \sqrt {-g}H_{MNL}A^{[M}F^{NL]}
  \eea
Choosing the gauge condition $B_{4 \mu} = 0$  and using the explicit form of the RS metric, we find
\bea
 S_{int}= \frac1 {3M_p^{\frac 3 2}} \int d^4 x \int d{\phi} \left[e^{2 \sigma}r_c
\eta^{\mu\alpha}\eta^{\nu\beta} \eta^{\lambda\gamma}
H_{\mu\nu\lambda}A_{[\alpha}F_{\beta\gamma]}
+\frac 6 {r_c}\eta^{\mu\alpha}\eta^{\nu\beta}(\partial_{\phi}B_{\mu\nu})A_{\beta}
(\partial_{\phi} A_{\alpha})\right] \eea 
Now using the Kaluza-Klein
decomposition for both the fields as mentioned
earlier,one obtains
\ber
 S_{int}&=& \frac 1 {3M_p^{\frac 3 2 }} \int d^4x \int d{\phi}
[e^{2 \sigma}
 \frac 1 {\sqrt {r_c}} \sum^{\infty}_{n,m,l=0}\xi^n \chi^m \chi^l \eta^{\mu\alpha}\eta^{\nu\beta}\eta^{\lambda\gamma}
H^n_{\mu\nu\lambda}A^m_{[\alpha}F^l_{\beta\gamma]}\nonumber\\
& &+ \frac 6 {r_c^{\frac 5 2}} \sum^{\infty}_{n,m,l=0}(\partial_{\phi}
\xi^n)\chi^m (\partial_{\phi}\chi^l)\eta^{\mu\alpha}\eta^{\nu\beta}
 B^n_{\mu\nu}A^m_{\alpha}A^l_{\beta}]\nonumber\\
 & & {}
 \eer
 Using eqn.(38) and (39), the part of the above action containing the massless modes only is given as 
\begin{equation}
S_{int}= \frac 1 {3M_p^{\frac 3 2}} \int d^4x \int d{\phi}\frac {e^{2 \sigma}} {\sqrt
{r_c}}\xi^0(\chi^0)^2 H_{\mu\nu\lambda}A^{[\mu}F^{\nu\lambda]}
\end{equation}
Where $\xi^0= \sqrt {k r_c}e^{-k r_c \pi}$ , $\chi^0= \frac 1 {2 \pi}$ and
$A^{\mu}=\eta^{\mu\nu}A_{\nu}$.\\
So the KR-EM part of the 4d effective action (without the curvature term) becomes 
\begin{equation}
S_{eff}= \int d^4x \left[- \frac 1 4 F^{\mu\nu}F_{\mu\nu} + \frac 1 2
H^{\mu\nu\lambda} H_{\mu\nu\lambda} + \frac C 3
H^{\mu\nu\lambda}A_{[\mu}F_{\nu\lambda]}\right]
\end{equation}
where $C = \sqrt{\frac k {M_p^3}} e^{k r_c \pi}$.\\
We have thus explicitly determined the KR-Maxwell coupling in the proposed string inspired RS scenario.\\  
By varying the $B_{\mu\nu}$ and $A_{\mu}$ as done in the previous section, we get the corresponding 
field equation for $B_{\mu\nu}$ and the set of 
modified Maxwell's equations.
From these equations the angle of optical rotation in this case turns out to be 
\begin{equation}
(\Delta \phi)_{mag} = -\sqrt{\frac {2k} {M_p^3}} e^{k r_c \pi} h \eta
\end{equation}
Once again we find that there is an enormous enhancement in the optical rotation by a large exponential warp factor.
We therefore find that for both the cases of bulk gauge field as well as the gauge field confined 
in the visible brane , a plane polarized electromagnetic wave will suffer an enormous optical rotation if our 
four dimensional world is an effective picture of a five dimensional Randall-Sundrum brane world. Interestingly
quite in contrary to our expectation, 
this enhancement takes place inspite of the fact that the massless mode of the KR field suffers large suppression
on the visible brane as shown in ref.\cite{bmssg}.
No such large rotation has ever been reported in any astrophysical experiments.
To make this result consistent with experiment one needs to finetune the value of h to an extremely small value 
which would render the theory unnatural.
This in a sense will bring back the old naturalness problem that we already have in connection with the 
stabilization of the Higg's mass and as a remedy of which the Randall-Sundrum scenario was originally proposed.\\
We thus conclude that if one believes in a string inspired 
cosmological model where the second rank antisymmetric tensor field is essentially present in the background 
then the $U(1)$ gauge anomaly cancelling Chern-Simons term results into a coupling 
between the KR and the electromagnetic field leading to optical rotation of plane of polarization of 
the distant galactic radiation.
However if one simultaneously 
considers a Randall-Sundrum type of brane world model to compactify one extra dimension then it is hard to explain 
the apparent anomaly between the 
theoretically predicted large value of the $\it {wavelength}$ ${independent}$ optical rotation and the corresponding 
small experimental value of this rotation measured in the context of distant galactic radio waves \cite{nodland}. 
The implications of the results reported in this work may also be studied in the context of observed CMB anisotropy.

One may try to explore the effect of the radion field on the KR field induced optical rotation for a possible suppression
in it's value.
It is also well known that the breakdown of supersymmetry may result into generation of mass for the scalar axion which in turn
may make their experimental signature unobservable. However among large class of possible vacua of string theory 
we still have not
been able to resolve that how and why a particular vacuum is chosen and the supersymmetry is broken. 
Nature of the scalar axion potential does depend on the geometry of the compact manifold as well as on the
exact mechanism of supersymmetry breaking.
Generation of the axion mass is therefore still not well understood.
The result reported in this work may however serve as an indirect clue to this problem by constraining the compactification scheme 
as well as  
supersymmetry breaking so that it may result into a significant axion mass and thereby making the resulting optical rotation
practically invisible from such a large value to make it consistent with observation. So far there is no proper understanding 
in this direction in the domain of string thoery.   
However in this work we find an undeniable conflict between the RS braneworld picture and the presence of
antisymmetric tensor field in the background spacetime. 

One of us (SSG) thanks I.Antoniadis, F.W.Hehl, M.Kamionkowski and P.Majumdar for their valuable comments and stimulating 
correspondences.DM acknowledges the Council for Scientific and Industrial Research, Government of India for 
providing financial support.

\end{document}